
\input phyzzx
\PHYSREV
\line{\hfill CU-TP-635 }
\line{ \hfill hep-th/9406049}
\FRONTPAGE
\strut\hfill {June 1994}
\vglue 1in
\font\twelvebf=cmbx12
\vskip 0.1in

\centerline{\twelvebf  Self-Dual SU(3) Chern-Simons Higgs Systems$^\dagger$}

\vskip .6in
\centerline{\it Hsien-Chung Kao and Kimyeong Lee }
\vskip .05in
\centerline{ Physics Department, Columbia University}
\centerline{ New York, N.Y. 10027, U.S.A.}
\vskip 0.8in
\centerline{\bf ABSTRACT}
\vskip 0.3in

We explore self-dual Chern-Simons Higgs systems with the local $SU(3)$
and global $U(1)$ symmetries where the matter field lies in the
adjoint representation. We show that there are three degenerate vacua
of different symmetries and study the unbroken symmetry and particle
spectrum in each vacuum.  We classify the self-dual configurations
into three types and study their properties.

\footnote{}{$\dagger$ This work is supported in part by the NSF
Presidential Young Investigator program (K.L.), the Alfred P. Sloan
Foundation (K.L.) and Department of Energy (H.-C.K.).}

\vfill
\endpage


\def\pr#1#2#3{Phys. Rev. {\bf D#1}, #2 (19#3)}
\def\prl#1#2#3{Phys. Rev. Lett. {\bf #1}, #2 (19#3)}

\def\np#1#2#3{Nucl. Phys. {\bf B#1}, #2 (19#3)}
\def\pl#1#2#3{Phys. Lett. {\bf B#1}, #2 (19#3)}
\def\tr{\,{\rm  tr}}
\def\ibid#1#2#3{{\it ibid}. {\bf #1}, #2 (19#3)}


\REF\rHong{J. Hong, Y. Kim, and P.Y. Pac,
       \prl{64}{2330}{90}; R. Jackiw and E.J. Weinberg,
       \ibid{64}{2334}{90}.}
\REF\rJackiw{ R. Jackiw, K. Lee, and E. Weinberg, \pr{42}{3488}{90}; Y.
Kim and K. Lee, \ibid{49}{2041}{94}. }

\REF\rKlee{K. Lee, \pl{255}{381}{91}; K. Lee \prl{66}{553}{91}; G.V. Dunne,
\pl{324}{359}{94}.}
\REF\rDunne{B. Grossman, \prl{65}{3230}{90}; G. Dunne, R. Jackiw, S.-Y.
          Pi and C. Trugenberger, \pr{43}{1332}{91};
          G. Dunne, Ann. Phys. {\bf 223}, 180 (1993); G.V. Dunne,
``Classification of Nonabelian Chern-Simons Vortices,'' Connecticut U.
Preprint UCONN-93-8}

\REF\rWitten{E. Witten and D. Olive, \pl{78}{97}{78}; P. Di Vecchia
and S. Ferrara, \np{130}{93}{77};  Z. Hlousek and D. Spector,
\ibid{397}{173}{93}.}

\REF\rClee{ C. Lee, K. Lee, and E.J. Weinberg, \pl{243}{105}{90};
 E.A. Ivanov, \ibid{268}{203}{91}; S.J. Gates and H. Nishino,
\ibid{281}{72}{92}; B.H. Lee, C. Lee and H. Min, \pr{45}{4588}{92}.}

\REF\rKao{H.-C. Kao and K. Lee, \pr{46}{4691}{92}; H.-C. Kao,
``Self-Dual Yang-Mills Chern-Simons Higgs Systems with an N=3 Extended
Supersymmetry,'' Columbia Univ. Preprint, CU-TP-595 (1993), to appear
in Phys. Rev. D. }

\REF\rSchonfeld{ J. Schonfeld, \np{185}{157}{81}; S. Deser, R. Jackiw
and S. Templeton, Ann. Phys. (N.Y.) {\bf 140}, 37 (1982).}

\REF\rKleee{ K. Lee, \np{373}{735}{92}; F.A. Bais, P. van Direl and M.
de Wild Propitius, \ibid{393}{547}{93}.}

\chapter{ Introduction}

There have been some recent activities related to the various
self-dual Chern-Simons Higgs systems. A bound on the energy functional
of these systems is saturated by the solitonic configurations of the
fractional spin. When the gauge symmetry is abelian, the structure of
these configurations has been studied quite
well.\refmark{\rHong,\rJackiw} While the self-dual systems with an
arbitrary gauge group and matter are shown to exist in the theories
with a global $U(1)$ symmetry, the self-dual configurations in these
systems have been studied only in the cases where either the gauge
symmetry or the matter is simple.\refmark\rKlee The nonrelativistic
limit of these system with matter in the adjoint representation has
been also studied extensively, where the classification of finite
energy soliton solutions has been found.\refmark\rDunne

 There are some studies of the simpler cases where the self-dual
solitons is equivalent to the abelian self-dual
solitons.\refmark\rKlee However, the soliton structure with nonabelian
symmetries turns out in general very intricate and interesting. In
this paper we study the theory with $SU(3)$ gauge group with matter
made of a complex scalar field in the adjoint representation.  This
model is one of the simple models with a nontrivial nonabelian feature
and exhibits rich vacuum and soliton structures. Solitons in this
theory would carry fractional spins and nonabelian charges. We hope
our work would shed some light on the general structures of the
self-dual systems.

First, we investigate the general consequences of the self-dual
equations for the configurations saturating the energy bound.  Then,
we show that there are three degenerate vacua of various unbroken
symmetries and topologies, and analyze the particles spectrum at each
vacuum. After that we study the characteristics of the self-dual
configurations and classify them into three types. In general we
expect these self-dual configurations describe topological and
nontopological solitons dwelling on each phase. We study the topology
of these solutions. Our analysis here provides a significant but not
complete understanding of these classical solutions of the self-dual
equations.

There is usually an underlying $N=2$ supersymmetry behind every
self-dual model.\refmark\rWitten There have been some studies of the
underlying $N=2$ supersymmetry in (Maxwell) Chern-Simons Higgs
systems.\refmark\rClee In addition, it is obvious that the maximum
possible symmetry for three dimensions is $N=3$ because a maximum
vector multiplet can have spin $1,1/2,0,-1/2$ up to sign. All $N=3$
supersymmetric theories have been constructed recently.\refmark\rKao

In Sec.~2, we review briefly the self-dual model with $SU(3)$
gauge group and a  complex scalar field in the adjoint representation.
We then investigate in some detail the restrictions on the field
configurations imposed by the self-dual
equations.  In Sec.~3 we
study the ground states, their symmetric properties and elementary
excitations.  In Sec.~4, we classify the  self-dual
configurations to three types and study their properties.  In Sec.~5
we conclude with some remarks.

\vskip 1.in

\chapter{ Model}

Let us consider a Chern-Simons-Higgs theory with  local $SU(3)$ and
global $U(1)$ symmetries.  The generators of $SU(3)$ in the
fundamental representation are made of $3\times 3$ Hermitian matrices,
$T^a$, with $a=1,2,,,8$ and satisfy the commutation relations
$[T^a,T^b] = if^{abc}T^c$ with $f^{abc}$ as the structure constants of
$SU(3)$. The normalization is such that $\tr T^aT^b =
\delta^{ab}/2$. The scalar matter field $\phi = (\phi_R^a + i\phi_I^a)T^a$
is made of a pair $\phi_R^a, \phi_I^a$ in the adjoint  representation of
$SU(3)$.  The Lagrangian density for the theory is given by
$$ {\cal L} = \kappa \epsilon^{\mu\nu\rho}\tr (A_\mu\partial_\nu
A_\rho - {2i\over 3}A_\mu A_\nu A_\rho ) +2\tr |D_\mu \phi|^2 -
{2\over \kappa^2} \tr \bigl| [\phi,[\phi^\dagger,\phi]] - v^2\phi \bigr|^2
\eqno\eq $$
where $D_\mu \phi = \partial_\mu \phi - i[A_\mu,\phi]$ with $A_\mu =
A^a_\mu T^a$. The gauge field strength is given by $F_{\mu\nu}\equiv
F_{\mu\nu}^a T^a = \partial_\mu A_\nu-\partial_\nu A_\mu -i
[A_\mu,A_\nu]$.  The theory is renormalizable because in $2+1$
dimensions the sixth order term in the potential energy has a
dimensionless coupling constant.  For the consistent quantum
mechanics, the coefficient $\kappa$ should be quantized so that
$\kappa = k/4\pi$ with a nonzero integer $k$.\refmark\rSchonfeld

Gauss's law constraint obtained from the variation of $A^a_0$ is
$$ {\cal G}  \equiv - \kappa F_{12} - i([D_0 \phi^\dagger, \phi]-
[\phi^\dagger, D_0\phi] ) = 0  \eqno\eq$$
The local gauge transformation generators are made of $\tr ({\cal G}T^a)$.
The Lagrangian (2.1) is invariant under the global phase rotation of
the scalar field. The  charge  density of the corresponding  global
$U(1)$ symmetry  is
$$  \rho_Q =2 i\tr (D_0\phi^\dagger\phi - \phi^\dagger D_0\phi)
 \eqno\eq$$
and the charge is  $Q = \int d^2r \rho_Q$.

We are interested in finding bound on the energy functional.  The
energy functional for the Lagrangian (2.1) is
$$ E = \int d^2r \,\, 2 \biggl\{  \tr |D_0\phi|^2 + {\rm tr}|D_i \phi|^2
+ {1\over \kappa^2} \tr |[\phi,[\phi^\dagger, \phi]] - v^2\phi|^2
\bigg\} \eqno\eq$$
With Gauss's law (2.2), the second term in the bracket becomes
$$ \eqalign{  \tr |D_i\phi|^2  =  &\ \tr |(D_1\pm iD_2)\phi|^2 \cr
&\  \pm {i\over \kappa}
\left\{ \tr \bigg[ D_0\phi^\dagger ( [\phi,[\phi^\dagger,\phi] -
v^2\phi) \bigg]  -h.c. \right\}
\pm {v^2 \over 2 \kappa} \rho_Q \cr }
\eqno\eq $$
up to a total derivative. This allows us to put $|D_0\phi|^2$ and the
potential energy density into a total square.  After integrating by
parts, the energy functional can be written as
$$ \eqalign{ E = &\ \int d^2r \,\, 2 \biggl\{ {\rm tr}\left|D_0\phi \pm
{i \over \kappa } \bigl(
[\phi,[\phi^\dagger,\phi]]  - v^2\phi \bigr ) \right|^2
 + {\rm tr} \left|D_1\phi \pm iD_2 \phi \right|^2 \biggr\} \cr
&\  \pm { v^2\over
\kappa } Q  \cr} \eqno\eq$$

Since the integrand in Eq.(2.6) is nonnegative, there  is a bound on
the energy functional
$$ E \ge m|Q|   \eqno\eq $$
where $m \equiv v^2/\kappa$ is the mass of elementary particles in
the symmetric  phase.
The bound is given by a total global charge, which is not a priori
related to any topological quantity.  The bound (2.7) is saturated by
the configurations satisfying Gauss's law and the  self-dual equations
$$ D_0\phi \pm {i\over \kappa }\bigl( [\phi,[\phi^\dagger,\phi]] -
v^2\phi \bigr) =0
\eqno\eq$$
$$ D_1\phi \pm iD_2\phi = 0  \eqno\eq$$
where the upper (lower) sign corresponds to the positive (negative) value
of $Q$.  Gauss's law and Eq.(2.8) can be combined to
$$  F_{12} = \pm   {2\over \kappa^2 }
([\phi^\dagger,[\phi,[\phi^\dagger,\phi]]] - v^2 [\phi^\dagger,\phi])
 \eqno\eq $$

When Eq.(2.8) is satisfied the $U(1)$ charge density (2.3) becomes
$$\rho_Q = \mp {4\over \kappa}(\tr[\phi^\dagger,\phi]^2
-v^2\tr\phi^\dagger\phi) \eqno\eq $$
For the configurations satisfying Eqs.(2.8) and (2.9), the total angular
momentum becomes
$$ \eqalign{ J = &\  -\int d^2r \,\, 2 \epsilon_{ij} r_i \,\, \tr
\bigl(
D_0\phi^\dagger D_j \phi + D_j\phi^\dagger D_0\phi \bigr) \cr
= &\  \int d^2r \,\,  r_i \partial_i  \,\, \tr \bigl(
[\phi^\dagger,\phi]^2 - 2v^2
\phi^\dagger \phi \bigr) \cr
= &\   2 \int d^2r \left\{  v^4 C_{\rm phase}-  \tr \biggl(
[\phi^\dagger,\phi]^2 - 2v^2\phi^\dagger \phi  \biggr)
  \right\} \cr}
\eqno\eq $$
where nonnegative $C_{\rm phase}$ is the spatial asymptotic value of
$\tr( [\phi^\dagger,\phi]^2 - 2v^2 \phi^\dagger \phi)/v^4$.  $C_{\rm
phase}$ depends on the phase or vacuum the system resides and will be
calculated Sec.3 for each phase.  As the potential energy density is
nonnegative, the ground states are characterized by the zeros of the
potential energy density, satisfying
$$ [\phi,[\phi^\dagger,\phi]] - v^2\phi = 0 \eqno\eq $$
which implies $\tr\phi^n = 0$ for any natural number $n$.

Let us now explore some aspects of the self-duality equations
Eqs.(2.9) and (2.10).  One can easily see that the solutions of these
equations satisfy $\partial_0 \tr \phi^n \mp {n\over
\kappa}v^2\tr\phi^n = 0 $ and $(\partial_1\pm i\partial_2)\tr \phi^n =
0$ for any natural number $n$, which imply that $\tr \phi^n$ is
a (anti)holonomic function.  As the field configuration approaches one
of vacua at spatial infinity where  $\tr \phi^n = 0$ because of
Eq.(2.13),  the holonomic function should vanish everywhere, i.e.,
$\tr \phi^n(x)=0$.  After triangularization with a similar
transformation, the trace conditions imply that the diagonal elements
of the triangularized matrix vanish, leading to  $\phi^n=0$
for $n \ge 3$.

The relation $ \phi^3(x) = 0$ everywhere is an important property
of the self-dual configurations.  If $\phi^3=0 $ and $\phi^2\neq 0$, there
is a three dimensional complex vector $\vec{u}$ such that $\phi^2
\vec{u} \neq 0$. Three vectors $\vec{u}, \phi\vec{u},
\phi^2\vec{u}$ are linearly independent and form a basis of a three
dimensional complex vector space.  Starting from $\phi^2\vec{u}$, we can
form an orthonormal basis by the standard procedure in linear algebra.
In this orthonormal basis, $\phi$ becomes
$$ \phi = v \left(\matrix{0 & f & h \cr
                        0 & 0 & g \cr
                        0 & 0 & 0 \cr}\right) \eqno\eq $$
where dimensionless $f,g,h$ are in general complex.  If $\phi^2=0$ and
$\phi\neq 0$ then, one can show easily that $\phi$ is again given by
Eq.(2.14) with either $f$ or $g$ to be zero. As there is an
orthonormal basis where $\phi$ is given by the above triangular
matrix, one can see that there is always a special unitary
transformation from any $\phi$ satisfying $\phi^3=0$ to this
triangular matrix.  We can use a local gauge transformation to put the
$\phi$ field in the above form at each spacetime point. Here we will
not consider the possibility of configurations, e.g., magnetic
monopole instantons, for which there may be a topological obstruction
to choose such a gauge globally.

\vskip 1.1in

\chapter{ Ground States and Spectra }

Let us now  consider the ground states of the model and explore the
unbroken symmetries and particle spectra.  As the potential energy in
the Lagrangian (2.1) is nonnegative, the ground state configurations
of zero energy satisfy Eq.(2.13). By identifying $J_z = [\phi^\dagger,
\phi]/v^2$, $J_+ = \sqrt{2}\phi^\dagger/v$, $J_- = \sqrt{2}\phi/v$,
one can see that $J_x = (J_+ + J_-)/2$, $J_y= (J_+ - J_-)/2i$ and
$J_z$ satisfy the angular momentum commutation relation. As $\phi$ is
a $3\times 3$ triangular matrix, `the total angular momentum' can be
zero, one half and one, having one, two and three dimensional
representations, respectively.

Alternatively, we notice that the vacuum configurations are the
solutions of the self-dual equations. Thus a vacuum configuration can
be chosen to the triangular form (2.14).  We solve Eq.(2.13) with the
triangular scalar field (2.14) to find the vacuum configurations.
Since the vacuum energies of three phases are degenerate, there will
be topological domain walls interpolating two different vacua.

\noindent{ 1) { \it  Phase I}}

Let us consider first the one-dimensional representation. The vacuum
expectation value of $\phi$ becomes $<\phi> = 0 $.  This is the
symmetric phase or {\it Phase I} where the global $U(1)$ and local
$SU(3)$ symmetries are preserved. There is no propagating mode for the
gauge field. The scalar field $\phi$ carries unit global charge and
forms the adjoint representation of $SU(3)$. The
mass of the scalar field is $m \equiv {v^2/\kappa}$.  $C_{\rm
phase}$ of Eq.(2.12) vanishes.

\noindent{ 2) {\it Phase II}}

For the two-dimensional representation the  vacuum
expectation value of $\phi$ can be chosen to be
$$ \vev{\phi}_v = {v\over \sqrt{2}} \left(\matrix{ 0 & 0 & 1 \cr
                              0 & 0 & 0  \cr
                             0 & 0 & 0 \cr}\right) \eqno\eq $$
In this phase $C_{\rm phase} $ of Eq.(2.12) becomes $1/2$.
In the unitary gauge, the scalar field becomes
$$
 \phi = {1\over \sqrt{2}}
            \left(\matrix{ \alpha/\sqrt{6}  &  0   & v +
		\delta/\sqrt{2}   \cr
                  \beta_1 & -2\alpha/\sqrt{6}  & 0  \cr
                  \gamma & \beta_2 & \alpha/ \sqrt{6} \cr}\right)
 \eqno\eq$$
All components except $\delta$ are complex. The fields are normalized
to have the standard kinetic term.   The masses of the
fields are $ m_\alpha =m$, $m_{\beta_1} = m_{\beta_2} = 3m/2 $, and
$m_\gamma=2m$, and $m_\delta=2m$.  The gauge field in the unitary
gauge becomes
$$
  A_\mu  =  {1\over \sqrt{2}}
      \left(\matrix{a_\mu /\sqrt{6}+b_\mu/\sqrt{2}  & c_{1\mu}   & d_\mu \cr
         \bar{c}_{1\mu} &-2a_\mu /\sqrt{6}  & c_{2\mu}  \cr
              \bar{d}_\mu & \bar{c}_{2\mu} & a_\mu /\sqrt{6}-b_\mu /\sqrt{2}
\cr}\right) \eqno\eq $$
with masses  $m_b  = 2m$, $m_{c_1}=m_{c_2} = m/2$, $m_d = m$.
The field $a_\mu$ is the gauge field for the leftover abelian  local gauge
symmetry and there is no corresponding propagating degrees of freedom.

By examining the symmetry generators which leave Eq.(3.1) invariant,
we can find the unbroken generators.  The $SU(3)\times U(1)$ group is
spontaneously broken to the global $U(1)_R$ and local $U(1)_S $
symmetry group. With the definition, $\tilde{T}^3 \equiv {\rm
diag}(1,0,-1)$ and $\tilde{T}^8 \equiv {\rm diag} (1,-2,1)/\sqrt{3}$,
the generators of the unbroken symmetries are given as $R = \int d^2r
( \rho_Q - \tr \tilde{T}^3 {\cal G} ) $ and $S = \int d^2r \tr
\tilde{T}^8{\cal G}$.

Since the gauge group acting on the adjoint representation
is really $SU(3)/Z_3$ where $Z_3$ is the center of $SU(3)$, the
vacuum manifold of {\it Phase II} would be $[SU(3)/Z_3 \times
U(1)_Q]/[U(1)_R \times U(1)_S]$ and $7$ dimensional.  We argue in the
next section  that the first fundamental homotopy
group of this vacuum manifold is $Z_2$.

We can write the lagrangian in terms of the fields (3.2) and (3.3),
which would be invariant  under these unbroken symmetries. One can
calculate the charge density for these generators.  For the global
$U(1)_R$ symmetry, the charge density is given by
$$ \eqalign{  \rho_Q- \tr {\cal G} \tilde{T}^3 =
&\ i\biggl\{ (\alpha \pi_\alpha- \bar{\alpha}\pi_{\bar{\alpha}}) +
{3\over 2}(\beta_i \pi_{\beta_i} - \bar{\beta}_i  \pi_{\bar{\beta}_i})
+ 2(\gamma \pi_\gamma - \bar{\gamma} \pi_{\bar{\gamma}}) \biggr\} \cr
&\ + {\kappa \over 2} \biggl\{ 2 \vec{\nabla}\times \vec{b} -i
\vec{c}_1\times\vec{\bar{c}}_1 -i \vec{c}_2\times \vec{\bar{c}}_2
-2i \vec{d}\times \vec{\bar{d}} \biggr\} \cr } \eqno\eq $$
For the local  $U(1)_S$ symmetry, the charge density is
$$ \eqalign{ \tr {\cal G}\tilde{T}^8 = &\ 3 \biggl\{ -( \beta_1 \pi_{\beta_1} -
\bar{\beta}_1 \pi_{\bar{\beta}_1})
+ (\beta_2 \pi_{\beta_2} - \bar{\beta}_2 \pi_{\bar{\beta}_2}) \biggr\} \cr
&\ + \kappa \biggl\{ \vec{\nabla}\times\vec{a} + 3 i\vec{c}_1\times
\vec{\bar{c}}_1 - 3i\vec{c}_2\times \vec{\bar{c}}_2 \biggr\}
\cr } \eqno\eq $$
In {\it Phase II}, the  energy bound (2.7) becomes
$$ {\cal E} \ge \pm m\int d^2r (\rho_Q - \tr \tilde{T}^3{\cal G} ) \eqno\eq $$
because of Gauss's law on the physical configurations. Note that all
particles of global $U(1)_R$ charge saturate the above bound.

Here we should note that there could be magnetic monopole instantons
in this phase. leading to the violation of the charge (3.5) for the
local gauge symmetry.\refmark\rKleee The gauge charge would be
conserved modulo a integer which depends on the coefficient of the
Chern-Simons term and the minimum monopole magnetic flux.  Further
investigation is necessary to settle this interesting possibility.

\noindent { 3) {\it  Phase III}

The three dimensional representation would lead to  the  ground
configuration
$$ \vev{\phi}_v = v \left(\matrix{ 0 & 1 & 0 \cr
                             0 & 0 & 1 \cr
                             0 & 0 & 0 \cr } \right) \eqno\eq $$
In this phase $C_{\rm phase} $ in Eq.(2.12) becomes $2$.
In the unitary gauge we can choose the scalar field to be
$$
 \phi = \left(\matrix{ \chi/\sqrt{30}  &  v+ (\zeta+ \xi)/2\sqrt{2}
     & -3\bar{\chi}/\sqrt{30} \cr
              \eta/2  & -2\chi/\sqrt{30}  & v+(\zeta-\xi)/2\sqrt{2}  \cr
               \psi/\sqrt{2} & \eta/2 & \chi/\sqrt{30} \cr}\right) \eqno\eq$$
The mass spectrum is given by $m_\chi = 5m$, $m_\eta =2m $, $m_\psi =
3m$, $m_\zeta = 2m$ and $m_\xi = 6m$.  The gauge field in the unitary
gauge is given by
$$
A_\mu=\left(\matrix{p_\mu/2 +q_\mu/2\sqrt{3}& (r_\mu + s_\mu)/2 &
t_\mu/\sqrt{2}  \cr
      (\bar{r}_\mu+\bar{s}_\mu)/2  & -q_\mu/\sqrt{3} & (r_\mu-s_\mu)/2   \cr
            \bar{t}_\mu/\sqrt{2} & (\bar{r}_\mu-\bar{s}_\mu)/2  &
 -p_\mu/2 + q_\mu/2\sqrt{3} \cr}\right) \eqno\eq $$
with masses $m_p = 2m$, $m_q = 6m$, $m_r=m$, $m_s=5m$, and $m_t=2m$.

The original symmetry is then spontaneously broken to
a global $U(1)_U$ symmetry. The generator of this symmetry is
$U = \int d^2r (\rho_Q - 2\tr \tilde{T}^3 {\cal G} ) $, where
$$ \eqalign{ \rho_Q - 2\tr{\cal G}\tilde{T}^3 = &\ i\biggl\{
(\chi \pi_\chi -\bar{\chi}\pi_{\bar{\chi}} ) + 2( \eta \pi_\eta -
\bar{\eta}\pi_{\bar{\eta}}) + 3(\psi \pi_{\psi} -
\bar{\psi}\pi_{\bar{\psi}})
 \biggr\} \cr
&\ + \kappa\biggl\{ 2 \vec{\nabla}\times \vec{p} +
i\vec{r}\times\vec{\bar{r}}
 +i\vec{s}\times\vec{\bar{s}}
+ 2i\vec{t}\times\vec{\bar{t}} \biggr\} \cr} \eqno\eq $$
In {\it Phase III}, the energy bound (2.7) becomes
$$ E\ge  \pm m      \int d^2r ( \rho_Q - 2\tr{\cal G}\tilde{T}^3 )
  \eqno\eq $$
due to Gauss's law.  The masses of all charged field except those of
$\chi$ and $s_\mu$ saturate the energy bound. Usually the masses of
charged particles in self-dual models saturate the energy bound. This
seems to be the first  example where the bound is  not saturated
by  charged particles. If our theory is a part of $N=2$
supersymmetric theory, $\chi$ and $s_\mu$ would be the bosonic part of
a vector supermultiplet.

The vacuum manifold of {\it Phase III} would be given by a
8-dimensional space  $[SU(3)/Z_3 \times U(1)_Q]/U(1)_U$,  We argue in
the next section that the first homotopy group of this manifold is
$Z_3$.

\endpage

\chapter{Self-Dual Configurations}

In this section, we study some properties of the self-dual
configurations which satisfy  Eqs.(2.9) and (2.10). Let us first try to
classify the possible configurations. There is always a gauge where
the scalar field is given by Eq.(2.14) as argued before.  For
convenience, we introduce three dimensionless quantities
$F,G,H$ such that
$$ F = |f|^2,\,\, G=|g|^2,\,\, H=|h|^2 \eqno\eq $$
where $f,g,h$ are given in Eq.(2.14).

{}From gauge invariant combinations $\tr \phi^\dagger \phi$, $\tr
\phi^\dagger \phi^2$, $\tr (\phi^\dagger\phi)^2$, and $ {\rm
det}[\phi^\dagger,\phi] $ of the scalar field, one can obtain some
dimensionless gauge invariant quantities,
$$ \eqalign{ \ & \ K = F +G + H \cr
 \ & \ L  = FG  \cr
 \ & \ M =  fg\bar{h} \cr
 \ & \ N = (F-G)L  \cr }  \eqno\eq $$
We classify the nontrivial ($K \neq 0$) solutions of the self-dual
equations into three types: {\it Type A} with $M=L=0$, {\it Type B}
with $M=0,L\neq 0$ and {\it Type C} with $M\neq 0$. As we will see,
{\it Type A} is the simplest and {\it Type C} is the most complicated
and interesting.  For each phase studied in the previous section, the
above three types of self-dual solutions might exist.
Some of them would be topological and others would be nontopological.

In terms of the above gauge invariant quantities, the global charge
density (2.11) for the self-dual configurations becomes
$$ \rho_Q = \pm  4mv^2 [  K(1-2K) +6 L ]
\eqno\eq $$
The total angular momentum (2.12) becomes
$$ J =  2 v^4 \int d^2r \left\{ C_{\rm phase}  +2 (K - K^2 + 3L ) \right\}
\eqno\eq $$

To understand further implications of the self-dual equations,  we define
$\partial = \partial_1 + i\partial_2$, $\bar{\partial}=
\partial_1 - i\partial_2 $, $A = A_1 + iA_2$ and $\bar{A} = A_1 -i
A_2 $. The magnetic field becomes $F_{12} = (\bar{\partial} A -
\partial \bar{A} - i[\bar{A}, A])/2i $. {}From now on we will be only
interested in  positive $Q$ configurations.
Eq.(2.9) can be written as
$$ \partial \phi - i [A,\phi] = 0 \eqno\eq $$
With $\phi$ given in  Eq.(2.14), the above equation implies that  $A$
should be an traceless triangular matrix,
$$ A = \left( \matrix{ a_1 &  b_1 &  c  \cr
                       0   & -(a_1+a_2)  & b_2 \cr
                       0   &  0   & a_2 \cr}\right) \eqno\eq $$
Furthermore  Eq.(4.5) in components becomes
$$ \eqalign{ \partial f - i(2a_1 + a_2) f = 0 \cr
             \partial g + i(a_1+ 2a_2) g = 0 \cr
             \partial h -i(a_1 - a_2) h + ib_2 f -ib_1 g =
0\cr}\eqno\eq $$

With the gauge field (4.6), the off-diagonal components of the
self-dual equation (2.10) become
$$ \eqalign{ \partial \bar{b}_1 + i(2a_1+a_2)\bar{b}_1 -ib_2\bar{c} -
4im^2 (2K -1)g\bar{h}=0 \cr
\partial\bar{b}_2 - i(a_1+2a_2)\bar{b}_2 + ib_1\bar{c} +
 4im^2 (2K -1)f\bar{h} = 0
\cr
\partial\bar{c} +i(a_1-a_2)\bar{c} = 0 \cr
 }\eqno\eq $$
The diagonal components  of Eq.(2.10) become
$$ \eqalign{ &\ \bar{\partial} a_1 - \partial \bar{a}_1 +
i|b_1|^2  + i|c|^2 +4im^2 [ (2K-1)(K - G) - 3L] =0 \cr
&\ \bar{\partial}a_2 -\partial \bar{a}_2 - i|b_2|^2  - i |c|^2
-4i m^2 [ (2K-1)(K-F) -3L ] =0 \cr} \eqno\eq $$
The self-dual equations in components are then given by Eqs.(4.7),
(4.8) and (4.9).  Let us now examine more closely what the self-dual
equations imply  for each type of solutions.

\noindent  {\it 1) \it Type A  Solutions }

For {\it Type A} solutions we can see easily that there is a local
gauge transformation where $f\neq 0$ and $g=h=0$ everywhere. (This is
gauge equivalent to the case only $h$ is not vanishing. )  Thus,
this type of configuration can exist only in {\it Phases I, II}.  As
there is no contribution to the energy from $a_2, b_2,c$, we can
regard $a_2,b_2,c$ to be zero. (In the lagrangian equation, the field
strength is zero and so the vector potential can be chosen to be
zero.)  Then, Eqs.(4.7) and (4.8) lead to
$$ \partial (f\bar{b}_1) = 0 \eqno\eq $$
As $f\bar{b}_1$ is holonomic function and the gauge field goes to zero
at the spatial infinity, $\bar{b}_1$ should be zero everywhere.

 Thus Eqs.(4.7) and (4.8) become  identical to the
self-dual equations studied before Ref.[1,2] with different numerical
factors:
$$ \eqalign{ &\ \partial f  -2i a_1 f = 0\cr
&\  {1\over 2i} (\bar{\partial}a_1 - \partial \bar{a}_1)
+2m^2 F(2F-1) = 0 \cr}\eqno\eq $$
With $\ln f \equiv {1\over 2} \ln F + i\sum_\alpha {\rm Arg}
(\vec{r}-\vec{q}_\alpha )$ with vortices at $\vec{q}_\alpha$, we can
combine the above two equations to
$$ \vec{\nabla}^2  \ln F + 8m^2 F(1- 2F) = 4\pi \sum_\alpha
\delta(\vec{r}-\vec{q}_\alpha)
\eqno\eq $$
where $\epsilon_{ij} \partial_i \partial_j {\rm Arg}(\vec{r} -\vec{q})
= 2\pi \delta (\vec{r} -\vec{q}) $ is used.  The solutions of this
equation are made of Q-balls in the symmetric {\it Phase I} and
vortices in the asymmetric {\it Phase II}.

The topology of vortices in the asymmetric phase is interesting.  In
the $SU(2)$ case, vortices are shown to have the $Z_2$
topology.\refmark\rKlee In our case, the $f$ field of an elementary
vortex in {\it Phase II} would be given as $f \approx
e^{i\varphi}/\sqrt{2}$ in large distance, which is equivalent to applying
a gauge transformation $\exp[ i \varphi ( \lambda_3/2 \pm
\lambda_8/\sqrt{3}) ]$ on the vacuum expectation value $f= 1/\sqrt{2}$.
Since both of these mappings lead to the same vortex and are
nontrivial elements of the first homotopy group $\pi_1(SU(3)/Z_3) =
Z_3$, the topology of elementary vortices in {\it Phase II} should be
$Z_2$.  This can be confirmed by noticing that the asymptotic
$e^{i2\varphi}/\sqrt{2}$ of the $f$ field for vorticity 2 is
represented by the gauge transformation $\exp[ i
\varphi \lambda_3]$ which is a trivial element of $\pi_1(SU(3)/Z_3)$.

There is a simplification of the global charge and the total angular
momentum.  Eqs.(4.3) and (4.11) leads to the charge density as a total
derivative
$$ \rho_Q =  2\kappa \vec{\nabla} \times \vec{a}_1
\eqno\eq $$
The total global charge would then get a contribution only from
spatial infinity. To simplify the angular momentum (2.12), we
introduce a transverse vector $\vec{\tilde{a}}_1 = \vec{a}_1 -(1/2)
\sum_\alpha \vec{\nabla} {\rm Arg} (\vec{r} - \vec{q}_\alpha) $, which
is not well defined at the vortex position.  Since the angular
momentum density is finite everywhere,  there is no finite  contribution from
the
vortex positions to the angular momentum (2.12) and the integration
region may be reduced from $R^2$ to $R^2_* = R^2 - \{\vec{q}_\alpha
\}$.  The angular momentum for {\it Type A} can then  be written as
$$ \eqalign{ J &\ = -8 \kappa m^2 \int_{R^2_*} d^2r \,\, \vec{r}\times
\vec{\tilde{a}}_1 (2F - 1)F \cr
&\ = 4\kappa \int_{R^2_*}  d^2r \,\, \vec{r}\times \vec{\tilde{a}} \vec{\nabla}
\times \vec{\tilde{a}}_1 \cr
&\ = 4\kappa \int_{R^2_*} d^2r \vec{\nabla}\cdot \left\{
 {1\over 2} \vec{r} \,\,\, \vec{\tilde{a}}_1^{\,\,2} - \vec{\tilde{a}}_1
\vec{r}\cdot \vec{\tilde{a}}_1 \right\} \cr }
\eqno\eq $$
For a given {\it Type A} self-dual configuration,  the angular
momentum  can be evaluated as the sum of the boundary contributions
from the vortex positions and spatial infinity. Eq.(4.14) was used
extensively in the second paper of Ref.[2] to study the vortex
dynamics. Especially, the statistical phase  of vortices is argued to
be originated from both the Aharonov and Bohm phase and the quantum
Magnus phase. Similar arguments would apply to our case under the study.

\noindent{\it  2)  Type B Solutions}

Let us here start by considering {\it Types B, C} in general terms.
For {\it Type B,  C} solutions, $fg\neq 0 $ and from Eqs.(4.7) and
(4.8) we get
$$ \eqalign{ &\ \partial(fg\bar{c}) = 0 \cr
 &\ \partial(f\bar{b}_1 + g\bar{b}_2) -i(fb_2 - gb_1)\bar{c} = 0 \cr}
\eqno\eq $$
As gauge fields vanish at the spatial infinity, the first part of
Eq.(4.15) implies that $fg\bar{c}=0$ everywhere, which in turn implies
$c=0$ everywhere. The second part of the above equation implies
$f\bar{b}_1 + g\bar{b}_2 = 0$, which can be satisfied by introducing a
new variable $u$ such that
$$b_1 =-i\bar{g}u,\,\, b_2= i\bar{f}u \eqno\eq $$
Eqs.(4.7) and (4.16)  lead  to an equation for the field $h$,
$$ \partial h -i(a_1-a_2)h - (F+G) u = 0 \eqno\eq $$

For {\it Type B} solutions where  $h=0$, the off-diagonal elements
$b_i$'s of the gauge field vanish everywhere as we can see from
Eqs.(4.16) and (4.17). Eqs.(4.7) and (4.9)  become
$$ \eqalign{ &\  \partial f -i(2a_1 + a_2) f = 0 \cr
              &\ \partial g + i(a_1 + 2a_2) g = 0 \cr
             &\ \bar{\partial} a_1 - \partial \bar{a}_1 + 4im^2( 2F -
                 G - 1) F = 0 \cr
              &\  \bar{\partial} a_2 - \partial \bar{a}_2  - 4im^2 ( 2G
                   - F - 1) G = 0 \cr  }
\eqno\eq $$
Eq.(4.18) is invariant under two U(1) gauge symmetries. We choose the
gauge so that
$$ \eqalign{ &\ \ln f = {1\over 2} \ln F + i \sum_\alpha {\rm Arg}
(\vec{r}- \vec{q}_{f\alpha })
 \cr
&\ \ln g = {1 \over 2} \ln G + i \sum_\beta {\rm
Arg}(\vec{r}-\vec{q}_{g\beta})
 \cr} \eqno\eq $$
where $\vec{q}_{f\alpha}, \vec{q}_{g\beta}$ are positions of $f,g$
vortices.  Then,  Eq.(4.18) can be written as
$$ \eqalign{ &\
\vec{\nabla}^2 \ln F  - 4m^2 ( 4F^2
-2G^2- FG  -2F + G) = 4\pi \sum_\alpha \delta(\vec{r}-\vec{q}_{f\alpha} )
  \cr
&\ \vec{\nabla}^2 \ln G -4m^2 ( -2F^2 + 4 G^2 - FG +F -2G) = 4\pi
\sum_\beta \delta(\vec{r}-\vec{q}_{g \beta})   \cr }\eqno\eq $$

We expect {\it Type B} solutions in all three phases. In {\it Phase
II}, one of $f$ or $g$ would take the vacuum expectation value
$1/\sqrt{2}$ at spatial infinity. By  similar argument for vortices
of {\it Type A} in {\it Phase II}, elementary vortices of {\it Type B}
in {\it Phase II} would have a topology $Z_2$.
In {\it Phase III}, vortices of {\it Type B} would have the $Z_3$
topology. To see this, we assume that at spatial infinity $f \approx
e^{i k \varphi}$ and $g \approx e^{i l \varphi}$ with integers $k,l$.
This is equivalent to a gauge transformation $\exp[ i\varphi {\rm
diag} ( 2k + l, -k +l , -k - 2l)/3 ]$ of $\vev{\phi}$, which is a $Z_3$ element
of
$SU(3)$.

{}From Eq.(4.18), the global charge density (4.3) for {\it Type B}
becomes
$$  \rho_Q = 2  \kappa  \vec{\nabla} \times ( \vec{a}_1 - \vec{a}_2)
\eqno\eq $$
Since the gauge fields should be smooth functions, the total charge will
get a contribution only from spatial infinity. To understand the
angular momentum better, let us define
$$ \eqalign{ &\ \tilde{a}_1 = a_1 - [ 2
\partial {\rm Arg}(f) + \partial {\rm Arg}(g)]/3 \cr
&\ \tilde{a}_2 =
a_2 + [ \partial {\rm Arg}(f) + 2\partial {\rm Arg}(g) ]/3 \cr}
\eqno\eq$$
{}From Eq.(4.18), one can see they are transverse vector fields.  Similar to
{\it
Type A}, we subtract the vortex positions from the integration
domain, $R^2_* = R^2 - \{\vec{q}_{f\alpha} , \vec{q}_{g\beta}\} $
without any change of the angular momentum. With Eqs. (2.12) and (4.16)
the angular momentum becomes
$$ \eqalign{ J &\ = 2\kappa \int_{R^2_*} d^2r \left\{ \vec{\nabla} \times
\vec{\tilde{a}}_1  ( 2 \vec{r} \times \vec{\tilde{a}}_1 + \vec{r}\times
\vec{\tilde{a}}_2 ) + \vec{\nabla}\times \vec{\tilde{a}}_2 ( 2 \vec{r}
\times \vec{\tilde{a}}_2 + \vec{r} \times \vec{\tilde{a}}_1)     \right\}
\cr
&\ =
2 \kappa \int_{R^2_*} d^2r \,\, \vec{\nabla}\cdot \left\{
\vec{r} \,\,\vec{\tilde{a}}_i^2 - 2 \vec{\tilde{a}}_i
(\vec{r}\cdot\vec{\tilde{a}}_i)  + \vec{r}\,\,(\vec{\tilde{a}}_1\cdot
\vec{\tilde{a}}_2) - \vec{\tilde{a}}_1
(\vec{r}\cdot\vec{\tilde{a}}_2)-
 \vec{\tilde{a}}_2 (\vec{r}\cdot\vec{\tilde{a}}_1) \right\}
\cr}
 \eqno\eq $$
Thus, the total angular momentum would get  contributions from the
vortex positions and spatial infinity. For a given self-dual
configurations, we can write dowm the total angular momentum as a
function of vortex positions in principle. As discussed for {\it Type
A}, Eq.(4.23) would lead to a considerable understanding of the
dynamics of the slowly moving vortices.

\noindent{ 3){ \it    Type C Solutions}}

{}From Eqs.  (4.8) and (4.16),  we get the equation for the $u$ field,
$$  \partial \bar{u} + i (a_1-a_2)\bar{u} -4 m^2 (2K-1) \bar{h} = 0
 \eqno\eq$$
{}From Eqs.(4.7), (4.17) and (4.24),  we get
$$  \eqalign{ &\ \partial(fg \bar{u}) - 4 m^2 (2K -1) fg\bar{h} = 0 \cr
&\  \partial( h\bar{u}) - (F+G) |u|^2 -4m^2 (2K-1) H = 0  \cr}
\eqno\eq $$
which implies   that  $\partial(h\bar{u}) $ is a real field.

Since $M = fg\bar{h}$ is a gauge invariant quantity, the vorticities
of the $f$ and $g$ field should be closely related to that of the $h$
field.  However, it is not easy to see what kind of solutions will
exist because the self-dual equations Eqs.(4.7), (4.9), (4.17) and
(4.25) are rather complicated.  In principle, {\it Type C} could exist
in all phases of the theory. The topology of {\it Type C} vortices in
the broken phases would be identical to that of {\it Type B} vortices
because {\it Type C} solutions become {\it Type A} or {\it Type B}
solutions at spatial infinity.

Note that the charge density is given as a total derivative,
$$ \rho_Q  = {\kappa \over
2i}\left( \bar{\partial}(a_2-a_1) -\partial(\bar{a}_2-\bar{a}_1)
\right) + \kappa \partial(h\bar{u})
\eqno\eq $$
However,  we have not been successful to express the angular momentum
as a boundary contributions as in Eq.(4.23). The self-dual equations
satisfied by {\it Type C} is rather complicated and needs further
consideration.

\vskip 1.in

\chapter{Conclusion}

We have studied the self-dual Chern-Simons Higgs systems with $SU(3)$
gauge symmetry and $U(1)$ global symmetry. The matter field is made of
a complex scalar field in the adjoint representation.  Our work is a
first step towards understanding the self-dual Chern-Simons Higgs
systems where the nonabelian symmetry plays a crucial role.  We have
analyzed the vacuum structure, particle spectrum and unbroken
symmetries. In addition, we classified the self-dual configurations
into three types of increasing complexity. We have shown that vortices
in {\it Phase II} would have the $Z_2$ topology and vortices in {\it
Phase III} would have the $Z_3$ topology. We have seen the global
charge of the self-dual configurations is given as a boundary
contribution from spatial infinity, making topological the total
energy of those configurations.  In addition, the self-dual
configurations are characterized by the total angular momentum, which
we have shown to take a rather simple form for at least {\it Types
A,B}.

Ideally, we want to understand the nature of self-dual solitons
completely and there are many directions to take to reach that goal.
Here are some ideas to be explored: the rotationally symmetric
solutions, the topological domain walls interpolating degenerate
vacua, the self-dual solutions of {\it Type C}, the classical dynamics
of slowly moving solitons,  the relation between relativistic and
nonrelativistic solutions, and the possible magnetic monopole
instantons in {\it Phase II}.  We would like to understand the quantum
aspects of these solitons. One novel possibility might be the
``nonabelian Magnus force and phase'' between vortices in the
asymmetric phase.  We note that some of understandings gained here could be
easily generalized to the cases with more complicated gauge groups and
matter fields.

\vskip 1.in

\centerline{\bf Acknowledgement}

We thank Alexios Polychronakos for  useful discussions.

\vfill
\endpage

\refout

\end